\documentclass[12pt,onecolumn,draftclsnofoot]{IEEEtran}

\ifCLASSINFOpdf

\else

\fi

\usepackage[utf8]{inputenc}
\usepackage{amssymb}
\usepackage{stfloats}
\usepackage{placeins}

\usepackage{overpic}
\usepackage{caption}
\usepackage{amsmath}
\usepackage{comment}
\usepackage{mathtools}
\usepackage{booktabs}
\DeclarePairedDelimiter{\norm}{\lVert}{\rVert}
\usepackage{cite}
\usepackage{subfig}
\usepackage{lipsum}
\usepackage{graphicx}
\usepackage{textcomp,gensymb}

\usepackage{multicol}
\usepackage{algorithm}
\usepackage[noend]{algpseudocode}
\usepackage{tikz}
\usepackage[english]{babel}
\usepackage{amsthm}
\theoremstyle{plain}
\usepackage{dirtytalk}

\graphicspath{ {./images/} }

\usepackage{amsmath}

\usepackage{mathtools}
\DeclarePairedDelimiter\ceil{\lceil}{\rceil}

\usepackage{tabularx}
\newcolumntype{L}[1]{>{\raggedright\arraybackslash}p{#1}}
\newcolumntype{C}[1]{>{\centering\arraybackslash}p{#1}}
\newcolumntype{R}[1]{>{\raggedleft\arraybackslash}p{#1}}
\usepackage{multirow}
\usepackage{mathtools, cuted}

\addto\captionsenglish{}
\begin{document}

\title{Transmitter Side Beyond-Diagonal Reconfigurable Intelligent Surface for Massive MIMO Networks}

\author{\IEEEauthorblockN{Anup~Mishra, \IEEEmembership{Graduate Student Member, IEEE}, Yijie~Mao, \IEEEmembership{Member, IEEE}, Carmen~D'Andrea, \IEEEmembership{Member, IEEE}, Stefano~Buzzi, \IEEEmembership{Senior Member, IEEE} and Bruno~Clerckx, \IEEEmembership{Fellow, IEEE}\vspace{-0.5cm}}

\thanks{The authors A. Mishra and B. Clerckx are with the Department of Electrical and Electronic Engineering, Imperial College London, London SW7 2AZ,
UK. B. Clerckx is also with Silicon Austria Labs (SAL), Graz A-8010, Austria (e-mail: anup.mishra17@imperial.ac.uk; b.clerckx@imperial.ac.uk).}
\thanks{Y. Mao is with the School of Information Science and Technology, ShanghaiTech University, Shanghai 201210, China (e-mail: maoyj@shanghaitech.edu.cn).}
\thanks{The authors S. Buzzi and C. D'Andrea are with the Department of Electrical and Information
Engineering (DIEI) of the University of Cassino and Southern Latium,
03043 Cassino, Italy. S. Buzzi is also affiliated with Politecnico di Milano, 20133 Milan, Italy (e-mail: buzzi@unicas.it, carmen.dandrea@unicas.it)}}

\maketitle

\begin{abstract}
This letter focuses on a transmitter or base station (BS) side beyond-diagonal reflecting intelligent surface (BD-RIS) deployment strategy to enhance the spectral efficiency (SE) of a time-division-duplex massive multiple-input multiple-output (MaMIMO) network. In this strategy, the active antenna array utilizes a BD-RIS at the BS to serve multiple users in the downlink. Based on the knowledge of statistical channel state information (CSI), the BD-RIS coefficients matrix is optimized by  employing a novel manifold algorithm, and the power control coefficients are then optimized with the objective of maximizing the minimum SE. Through numerical results we illustrate the SE performance of the proposed transmission framework and compare it with that of a conventional MaMIMO transmission for different network settings.
\end{abstract}

\IEEEpeerreviewmaketitle
\vspace{-0.2cm}
\section{Introduction}\label{Intro}
\IEEEPARstart{T}{he} impeccable merits of massive multiple-input multiple-output (MaMIMO) have effectuated it as the backbone of modern day wireless networks\cite{massivemimobook,buzzi2022approaching}. The ability of MaMIMO to serve a large number of users in the same time-frequency resource block by aggressively multiplexing them in the spatial domain has led to unprecedented levels of network coverage and spectral efficiency (SE)\cite{massivemimobook}. This aggressive spatial multiplexing is enabled by leveraging joint coherent transmission/reception and a large number of fully digital radio frequency (RF) chains at the base station (BS). However, as the number of RF chains increases inordinately the total energy consumption increases linearly with it. But, the data rates only grow logarithmically at maximum\cite{buzzi2022approaching}. Such a situation is expensive and leads to energy inefficiency in the network. Therefore, there is a perpetual desire to find solutions that enhance the SE performance of a MaMIMO network for a small or moderate number of RF chains. 
\par One potential solution to boosting MaMIMO performance is utilizing a reconfigurable intelligent surface (RIS) to direct MaMIMO transmission\cite{RISTutorial@Zhang}. RISs, with their ability to shape the propagation channel by using passive elements, have gained significant traction as a promising candidate for efficiently enhancing the performance of a wireless network\cite{BSSide@Zhang,RISTutorial@Zhang}. In most existing works, RIS is typically deployed at the side of distributed users to minimize the infamous product-distance path loss\cite{BSSide@Zhang}. However, the same can be achieved by deploying RIS at the BS side as well. In fact, recent works have investigated BS side RIS deployment strategies and realized SE gains owing to their wider network coverage, higher passive beamforming gain, and lower RIS-BS signalling overhead\cite{BSSide@Zhang,original@arch_paper,buzzi2022approaching}. In regard to architecture, conventional RISs are postulated as diagonal RIS (D-RIS) architectures working under reflective mode and thus mathematically modelled as diagonal phase shift matrices\cite{RISTutorial@Zhang}. Alternatively, \cite{fullIRS@Shanpu} proposed a general beyond-diagonal RIS (BD-RIS) architecture using scattering parameter network analysis; categorized as single-, group-, and fully-connected RIS architectures; where the modelling goes beyond diagonal phase shift matrices. As opposed to D-RIS where no element is connected to the others, in BD-RIS, all or elements within a group are connected to each other. Moreover, BD-RIS architectures enable RIS to not only adjust the phase but also the magnitude of the impinging waves, thereby providing significant SE gains over D-RIS\cite{fullIRS@Shanpu}. 
\par Motivated by the aforementioned, in this letter we propose a BS side BD-RIS (fully-connected architecture) deployment with the aim of achieving an enhanced SE performance in a MaMIMO network. With the objective of making user channels `near-orthogonal', we formulate a novel manifold algorithm to optimize the BD-RIS coefficients matrix based on statistical channel state information (CSI). The power control coefficients are then optimized based on \textit{hardening bound} with the aim of maximizing the minimum SE among users. This is the first work that integrates BD-RIS with MaMIMO and optimizes the BD-RIS matrix using statistical CSI. Through numerical results, we first investigate and compare the SE performance of the proposed transmission framework with BD-RIS and D-RIS. Thereafter, we investigate different configurations of active antennas, passive RIS elements , and users for which the framework outperforms a conventional MaMIMO transmission.
 \begin{figure}
    \centering
    \subfloat[Pictorial representation of the considered BS side RIS deployment.]
    {\includegraphics[width=8cm,height=5.5cm, tics=10]{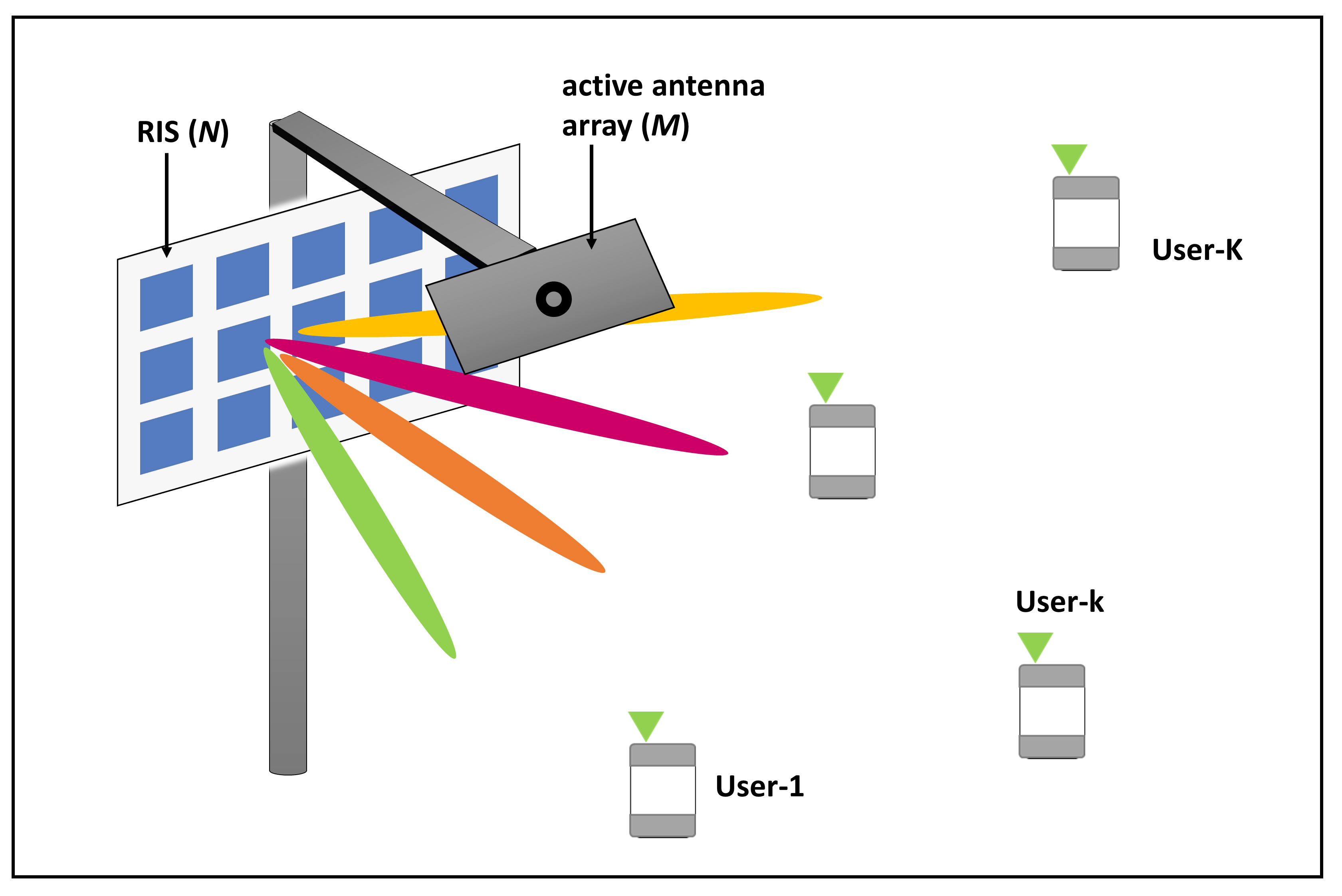}}\\
    \subfloat[An illustrative example of a $4$-element reconfigurable impedance
network.]
{\includegraphics[width=8cm,height=4cm]{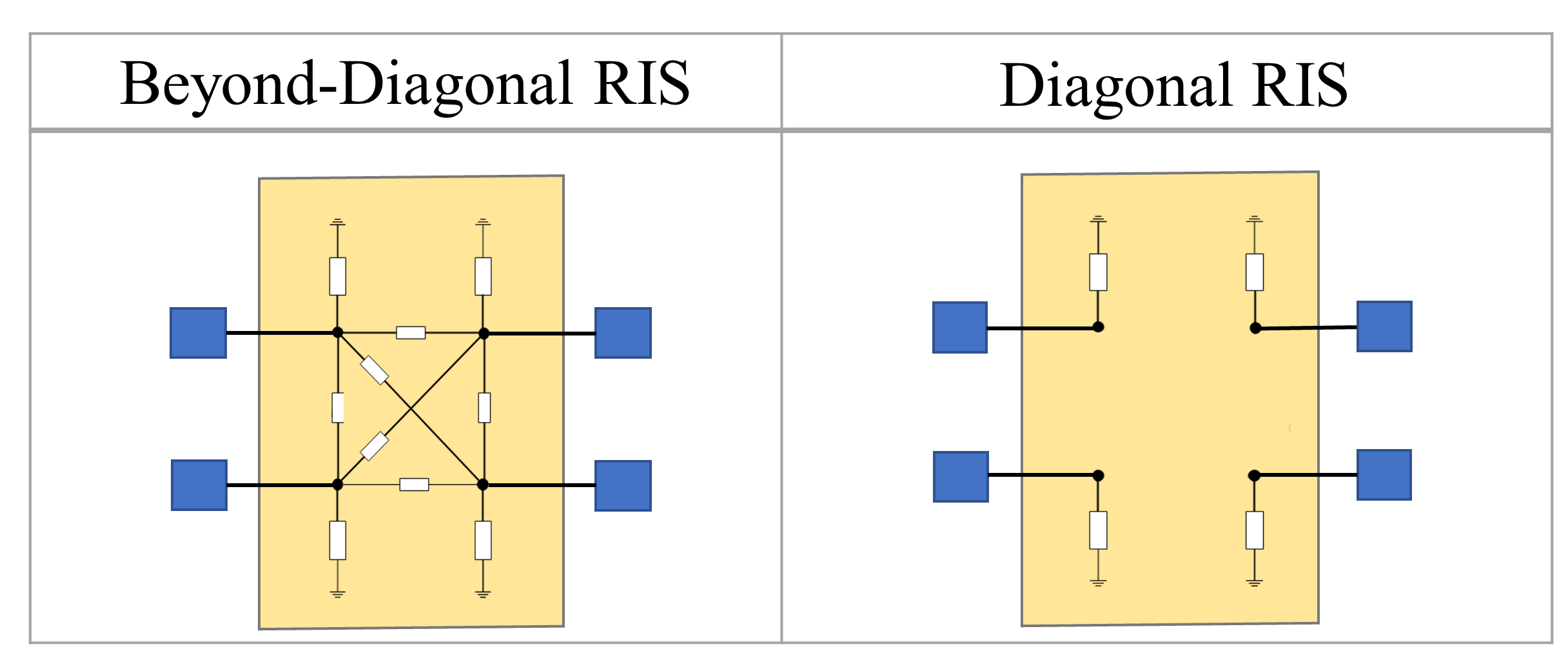}}\\
    \caption{BS side deployment (a) and RIS architectures (b).}%
    \label{fig:BS_Side_RIS}\vspace{-0.5cm}
\end{figure}
\vspace{-0.3cm}
\section{System Model}\label{Sysmod}
Consider a single-cell network with a BS side RIS deployment as depicted in Fig.~\ref{fig:BS_Side_RIS}(a), where the RIS is placed at a short distance (few wavelengths) from the active antennas and the BS serves multiple users via the RIS. We assume that the active antenna array-user link is significantly weaker than the RIS-user link and therefore can be neglected \cite{original@arch_paper,buzzi2022approaching}. Moreover, the blockage introduced by the active antenna array to RIS reflected signal is also neglected; which can be ensured by placing the active antenna array laterally with respect to the RIS. Further, the mutual coupling between active antenna elements or passive RIS elements is ignored by assuming inter-element distance greater than or equal to half wavelength for both. Finally, we assume that each RIS element is in the far field of each active antenna element. However, it is not required for the whole RIS to be in the far field of whole active antenna array\cite{buzzi2022approaching}. We request the readers to refer to \cite{buzzi2022approaching,original@arch_paper} for a detailed description of the considered RIS deployment. 
\par The BS is equipped with $M$ active antenna array  serving $K$ single-antenna users simultaneously via a RIS with $N>M$ passive elements. For generality, the active antenna elements and passive RIS elements both are placed in uniform planar arrays (UPAs), i.e., $N=N_{H}N_{V},\, N_{H}>\,N_{V}\geq1$ and $M=M_{H}M_{V},\, M_{H}>\,M_{V}\geq1$. The channel between the BS and RIS, denoted by $\mathbf{H}\in \mathbb{C}^{M\times N}$, can be written as  \cite{buzzi2022approaching}  
\begin{equation}
\mathbf{H}\left(i,j\right)=\sqrt{\varrho\,G_{A}\left(\theta_{i,j}\right)G_{R}\left(\theta_{i,j}\right)}\,\frac{\lambda}{4\pi\,d_{i,j}}\,e^{-i2\pi d_{i,j}/\lambda},
\end{equation}
where $\varrho\in \mathbb{R}$ models the reflection efficiency of the RIS, $G_{A}\left(\theta_{i,j}\right)$ and $G_{R}\left(\theta_{i,j}\right)$ represent the gains between the $i$-th active antenna element and $j$-th  RIS element corresponding to the look angle $\theta_{i,j}$, $d_{i,j}$ is the distance between the $i$-th active antenna element and $j$-th  RIS element, and $\lambda$ is the wavelength of the impinging wave on the RIS\cite{original@arch_paper,buzzi2022approaching}. 
\par In this work, we consider two different RIS configurations, namely, BD-RIS and D-RIS. Fig.~\ref{fig:BS_Side_RIS}(b) illustrates a $4$-element BD (left) and D (right) reconfigurable impedance network\cite{fullIRS@Shanpu}. In BD-RIS, each element of the reconfigurable impedance network is connected to other elements thereby generalizing the D-RIS in which none of the elements are connected to the others. Irrespective, the RIS reflection coefficients matrix can be modelled as $\boldsymbol{\Theta}\in \mathbb{C}^{N\times N}$ which can be reconfigured to adapt to the channel between the RIS and users\cite{fullIRS@Shanpu}. The channel between the RIS and user-$k$, denoted by $\mathbf{h}_{k}\in \mathbb{C}^{N}$, is considered to be spatially correlated and modelled as $\mathbf{h}_{k}=\sqrt{\mathbf{R}_{k}}\,\Bar{\mathbf{h}}_{k}$ such that $\mathbf{h}_{k}\sim \mathcal{CN}\left(\mathbf{0},\mathbf{R}_{k}\right)$.\footnote{Note that \cite{buzzi2022approaching} considers RIS-user channel fading to be uncorrelated assuming rich scattering at sub-$6$ GHz frequencies. However, we assume fading to be spatially correlated owing to a large number of RIS elements closely placed on a planar surface and directional nature of the propagation environment\cite{Kronecker@model,massivemimobook}.} Here,  $\mathbf{R}_{k}\in \mathbb{C}^{N\times N}$ is the spatial correlation matrix describing the large scale fading
property, accounting for both path loss and shadowing, and is characterized using the Kronecker model of \cite{Kronecker@model}. The small scale fading is captured by $\Bar{\mathbf{h}}_{k}\sim \mathcal{CN}\left(\mathbf{0},\mathbf{I}_{N}\right)$. Following, the composite channel between the BS active antenna array and user-$k$, denoted by $\mathbf{g}_{k}$, is written as 
\begin{equation}
\mathbf{g}_{k}=\mathbf{H}\boldsymbol{\Theta}\mathbf{h}_{k}.
\end{equation}
\par In regard to the transmission protocol, a standard time-division duplex (TDD) protocol is considered, where out of the $\tau$ available channel uses, $\tau_{up}$ are for uplink (UL) training phase, $\tau_{ud}$ for UL data transmission and $\tau_{d}$ for downlink (DL) data transmission. Naturally, $\tau\geq\tau_{up}+\tau_{ud}+\tau_{d}$. Here, we set $\tau_{ud}=0$ as we do not consider UL data transmission.
\vspace{-0.10cm}
\subsection{Channel estimation}
The BS side RIS architecture considered in our system model allows for the peculiar assumption that $\mathbf{H}$ is perfectly known to the BS\cite{original@arch_paper,buzzi2022approaching}. Therefore, the objective is to estimate the channel between the RIS and user-$k$. To this end, we consider that each user transmits a pilot sequence of length $\tau_{p}\leq\tau_{up}$. Denote the pilot sequence of user-$k$ as $\boldsymbol{\varphi}_{k}\in \mathbb{R}^{\tau_{p}}$ such that $\boldsymbol{\varphi}_{k}^{T}\boldsymbol{\varphi}_{k}=\tau_{p}$. Since $N>M$, estimating the $N$-dimensional channel vector $\mathbf{h}_{k}$ will require each user to transmit its pilot sequence $Q=\ceil{N/M}$ times, where each time (an epoch) the RIS configuration is different to guarantee that the number of unknowns are not greater than the number of observables\cite{buzzi2022approaching}. Next, with $\rho_{\textrm{uL}}$ as the UL transmit power available at each user, and $\boldsymbol{\Theta}_{\textrm{tr}}^{(q)}$ as the  configuration of the RIS in the $q$-th training epoch, the received signal at the BS, $\mathbf{Y}^{(q)}\in\mathbb{C}^{M}$, in the $q$-th training epoch is given by
\begin{equation}\label{eq:Rx_pilot}
\mathbf{Y}^{(q)}=\sum_{k=1}^{K}\sqrt{\rho_{\textrm{uL}}}\,\mathbf{H}\boldsymbol{\Theta}_{\textrm{tr}}^{(q)}\mathbf{h}_{k}\boldsymbol{\varphi}_{k}^{T} + \mathbf{N}^{(q)},
\end{equation}
where $\mathbf{N}^{(q)}\sim\mathcal{CN}\left(\mathbf{0},\sigma_{\textrm{uL}}^{2}\mathbf{I}_{M}\right)$ is the additive white noise. The estimate of channel $\mathbf{h}_{k}$ can be calculated by first projecting the received signal in \eqref{eq:Rx_pilot} along the pilot sequence of user-$k$ and obtaining $\mathbf{y}_{k,q}=\mathbf{Y}^{(q)}\boldsymbol{\varphi}_{k}$, given by
\begin{equation}
\mathbf{y}_{k,q}=\tau_{p}\sqrt{\rho_{\textrm{uL}}}\,\mathbf{g}_{k}^{(q)} + \sum_{i\neq k}^{K}\, \psi_{k,i}\, \sqrt{\rho_{\textrm{uL}}}\,\mathbf{g}_{i}^{(q)} + \mathbf{n}_{k,q},
\end{equation}
where $\mathbf{g}_{k}^{(q)}=\mathbf{H}\boldsymbol{\Theta}_{\textrm{tr}}^{(q)}\mathbf{h}_{k}$ is the composite channel of user-$k$ observed at the $q$-th epoch, $\psi_{k,i}=\boldsymbol{\varphi}_{k}^{T}\boldsymbol{\varphi}_{i}$, and $\mathbf{n}_{k,q}=\mathbf{N}^{(q)}\boldsymbol{\varphi}_{k}\sim \mathcal{CN}\left(\mathbf{0},\tau_{p}\sigma_{\textrm{uL}}^{2}\mathbf{I}_{M}\right)$. Thereafter, $\mathbf{y}_{k,q},\,\forall q\in \{1,\ldots,Q\}$ are stacked together to form the overall vector $\mathbf{y}_{k}=[\mathbf{y}_{k,1}^{T},\ldots,\mathbf{y}_{k,Q}^{T}]^{T}\in\mathbb{C}^{MQ}$, which can be written as
\begin{equation}\label{eq:Stacked_obs}
\mathbf{y}_{k}=\tau_{p}\sqrt{\rho_{\textrm{uL}}}\,\widetilde{\mathbf{H}}_{\textrm{tr}}\,\mathbf{h}_{k} + \sum_{i\neq k}^{K}\, \psi_{k,i}\, \sqrt{\rho_{\textrm{uL}}}\,\widetilde{\mathbf{H}}_{\textrm{tr}}\,\mathbf{h}_{i} + \mathbf{n}_{k},
\end{equation}
where $\mathbf{n}_{k}=[\mathbf{n}_{k,1}^{T},\ldots,\mathbf{n}_{k,Q}^{T}]^{T}\in\mathbb{C}^{MQ}$, and $\widetilde{\mathbf{H}}_{\textrm{tr}}\in\mathbb{C}^{MQ\times N}$ is a matrix defined as
\begin{equation}
\widetilde{\mathbf{H}}_{\textrm{tr}}=\Big[\left(\mathbf{H}\boldsymbol{\Theta}_{\textrm{tr}}^{(1)}\right)^{T},\ldots,\left(\mathbf{H}\boldsymbol{\Theta}_{\textrm{tr}}^{(Q)}\right)^{T}\Big]^{T}.
\end{equation}
{Finally, considering the singular value decomposition of $\widetilde{\mathbf{H}}_{\textrm{tr}}=\mathbf{U}\boldsymbol{\Lambda}\mathbf{V}^{H}$, and employing a conventional linear minimum-mean-square error (LMMSE) estimator, the estimate of $\mathbf{h}_{k}, \forall k$ can be obtained  as \cite{buzzi2022approaching,massivemimobook}
\begin{equation}
\widehat{\mathbf{h}}_{k}=\mathbf{V}\mathbf{R}_{\mathbf{v}_{k},\bar{\mathbf{y}}_{k}}\mathbf{R}_{\bar{\mathbf{y}}_{k},\bar{\mathbf{y}}_{k}}^{-1}\mathbf{U}^{H}\mathbf{y}_{k}\sim \mathcal{CN}\left(\mathbf{0},\mathbf{\Phi}_{k}\right),
\end{equation}
where $\mathbf{\Phi}_{k}=\mathbf{V}\mathbf{R}_{\mathbf{v}_{k},\bar{\mathbf{y}}_{k}}\mathbf{R}_{\bar{\mathbf{y}}_{k},\bar{\mathbf{y}}_{k}}^{-1}\mathbf{R}_{\mathbf{v}_{k},\bar{\mathbf{y}}_{k}}^{H}\mathbf{V}^{H}$ is the covariance matrix of the channel estimate of the link between the RIS and user-$k, \forall k$, with
\begin{subequations}\label{eq:Second_Stats}
\begin{align}
&\mathbf{R}_{\mathbf{v}_{k},\bar{\mathbf{y}}_{k}}=\tau_{p}\sqrt{\rho_{\textrm{uL}}}\;{\mathbf{V}}^{H}\mathbf{R}_{k}{\mathbf{V}}{\boldsymbol{\Lambda}}^{H},\\
&\mathbf{R}_{\bar{\mathbf{y}}_{k},\bar{\mathbf{y}}_{k}}=\left(\sum_{j=1}^{K}\psi_{jk}^{2}\,\rho_{\textrm{uL}}{\boldsymbol{\Lambda}}{\mathbf{V}}^{H}\mathbf{R}_{j}{\mathbf{V}}{\boldsymbol{\Lambda}}^{H}\right)+\sigma_{\textrm{uL}}^{2}\tau_{p}\,\mathbf{I}_{N}.
\end{align}
\end{subequations}
Further, the channel estimation error of the RIS-user link for user-$k$, $\widetilde{\mathbf{h}}_{k}=\mathbf{h}_{k}-\widehat{\mathbf{h}}_{k}$, is statistically independent from the estimates and is distributed as $\widetilde{\mathbf{h}}_{k}\sim \mathcal{CN}\left(\mathbf{0},\mathbf{R}_{k}-\mathbf{\Phi}_{k}\right),\,\forall k$.} 
\vspace{-0.2cm}
\subsection{Spectral Efficiency}
Denote the independently encoded stream of user-$k$ as ${s}_{k}\in\mathbb{C}$ such that $\mathbb{E}\{|s_{k}|^{2}\}=1, \forall k$. Further, denote $\rho_{k}$ as the transmit power assigned to user-$k$ such that $\sum_{k=1}^{K}\rho_{k}\leq \rho_{\textrm{dL}}$, where $\rho_{\textrm{dL}}$ is the total transmit power available at the BS. With $\mathbf{w}_{k}\in\mathbb{C}^{M}$ as the precoder for stream $s_{k}$, the signal transmitted at the BS is expressed as $\mathbf{x}=\sum_{k=1}^{K}\sqrt{\rho_{k}}\,\mathbf{w}_{k}\mathbf{s}_{k}$. We choose maximum ratio transmission (MRT) for precoding, given by 
\begin{equation}
\mathbf{w}_{k}=\frac{\mathbf{H}\boldsymbol{\Theta}\widehat{\mathbf{h}}_{k}}{\sqrt{\mathbb{E}\{\norm{\mathbf{H}\boldsymbol{\Theta}\widehat{\mathbf{h}}_{k}}^{2}\}}}=\frac{\mathbf{H}\boldsymbol{\Theta}\widehat{\mathbf{h}}_{k}}{\sqrt{\textrm{tr}\left(\mathbf{H}\boldsymbol{\Theta}\boldsymbol{\Phi}_{k}\boldsymbol{\Theta}^{H}\mathbf{H}^{H}\right)}}.
\end{equation}
Note that we consider the RIS matrix $\boldsymbol{\Theta}$ to be fixed over multiple coherence intervals, and thus, it can be configured only once before the DL data transmission commences.
\par In a conventional TDD MaMIMO protocol, since a user is only aware of the effective ergodic precoded channel, i.e., $\mathbb{E}\{\mathbf{g}_{k}^{H}\mathbf{w}_{k}\}$, the received signal at user-$k$ can be written as \cite{massivemimobook}
\begin{align}\label{eq:Rx_Signal}
    r_{k}&=\sqrt{\rho_{k}}\,\mathbb{E}\{\mathbf{g}_{k}^{H}\mathbf{w}_{k}\}s_{k}+\sqrt{\rho_{k}}\left(\mathbf{g}_{k}^{H}\mathbf{w}_{k}-\mathbb{E}\{\mathbf{g}_{k}^{H}\mathbf{w}_{k}\}\right)s_{k} +\sum_{i\neq k}^{K}\sqrt{\rho_{i}}\,\mathbf{g}_{k}^{H}\mathbf{w}_{i}s_{i}+z_{k},
\end{align}
where the first term in \eqref{eq:Rx_Signal} is the desired signal over the known `deterministic' channel, the second term is the interference caused due to the lack of complete channel knowledge at the user, the third term is the inter-user interference and $z_{k}\sim \mathcal{CN}\left(0,\sigma_{\textrm{dL}}^{2}\right)$ is the noise at user-$k, \forall k$ \cite{massivemimobook}. Given \eqref{eq:Rx_Signal}, and assuming that the power control coefficients are only dependent on channel statistics, the achievable signal to noise plus interference (SINR) of user-$k$ can be lower bounded as 
\begin{equation}\label{eq:Hardening_B}
    \gamma_{k}^{\textrm{LB}}=\frac{\rho_{k}|\mathbb{E}\{\mathbf{g}_{k}^{H}\mathbf{w}_{k}\}|^2}{\sum_{i= 1}^{K}{\rho_{i}}\,\mathbb{E}\{|\mathbf{g}_{k}^{H}\mathbf{w}_{i}|^2\}-{\rho_{k}}|\mathbb{E}\{\mathbf{g}_{k}^{H}\mathbf{w}_{k}\}|^2+\sigma_{\textrm{dL}}^{2}},
\end{equation}
also known as the \textit{hardening} \textit{bound}\cite{massivemimobook}. Consequently, the lower-bound of the SE of user-$k$ is defined as \cite{massivemimobook}
\begin{equation}\label{eq:SE_Exp}
    \textrm{SE}_{k}^{\textrm{LB}}=\left(1-\frac{\tau_{up}}{\tau}\right)\log_{2}\left(1+\gamma_{k}^{\textrm{LB}}\right),
\end{equation}
The expectations in \eqref{eq:Hardening_B} can either be computed in closed form or using Monte-Carlo simulations. Note that the hardening bound has been considered as a performance measure in a wide body of literature on TDD MaMIMO\cite{massivemimobook}. Therefore, we do not expound its derivation in this letter to avoid redundancy.
\section{Optimization Framework}\label{Opt_F}
In Section. \ref{Sysmod}, we define the precoders and SE expressions considering that both RIS matrix and power control coefficients are only dependent on the channel statistics. Such a consideration is much desired as the computational complexity of optimizing RIS matrix and power control coefficients would be exorbitantly high if done for every coherence interval. Therefore, in this section our aim is to formulate and optimize the aforementioned entities based on statistical CSI.
\vspace{-0.2cm}
\subsection{RIS optimization}
Since the RIS matrix can be conveniently configured to enhance the propagation environment, we optimize the RIS matrix with the objective of achieving `near-orthogonality' between composite channel vectors of different users\cite{buzzi2022approaching}. Taking the constraint of optimization based on statistical CSI into account, near-orthogonality is implied in an ergodic sense. Following, we formulate the cost function as\cite{buzzi2022approaching}   
\begin{align}\label{eq:Obj_RIS}
    f\left(\boldsymbol{\Theta}\right)=&\sum_{k=1}^{K-1}\sum_{j=k+1}^{K}\mathbb{E}\{|\mathbf{g}_{k}^{H}\mathbf{g}_{j}|^2\} =\,\sum_{k=1}^{K-1}\sum_{j=k+1}^{K}\textrm{tr}\left(\mathbf{R}_{k}\boldsymbol{\Theta}^{H}\mathbf{G}\boldsymbol{\Theta}\mathbf{R}_{j}\boldsymbol{\Theta}^{H}\mathbf{G}\boldsymbol{\Theta}\right),
\end{align}
with $\mathbf{G}=\mathbf{H}^{H}\mathbf{H}$. Since we consider a BD-RIS architecture proposed in \cite{fullIRS@Shanpu}, the RIS matrix optimization problem is formulated as\cite{fullIRS@Shanpu}
\begin{subequations}\label{eq:RIS_Opt_Prob}
\begin{align}
    \min_{\boldsymbol{\Theta}}&\;\; f\left(\boldsymbol{\Theta}\right) \\ &\;\;\boldsymbol{\Theta}^{H}\boldsymbol{\Theta}=\mathbf{I}_{N}\label{eq:const_mani}.
\end{align}
\end{subequations}
Constraint \eqref{eq:const_mani} makes problem \eqref{eq:RIS_Opt_Prob} challenging to solve. To that end, we employ manifold algorithm which allows us to construct all available solutions of the problem as a manifold and transform problem \eqref{eq:RIS_Opt_Prob} as an unconstrained optimization problem on that manifold\cite{boumal2023intromanifolds}.
{Here, constraint \eqref{eq:const_mani} forms a $N^{2}$ dimensional complex Stiefel manifold, i.e., $\mathcal{M}=\{ \boldsymbol{\Theta}\in \mathbb{C}^{N\times N}\mid \boldsymbol{\Theta}^{H}\boldsymbol{\Theta}=\mathbf{I}_{N}\}$, making problem \eqref{eq:RIS_Opt_Prob} an unconstrained optimization on $\mathcal{M}$, given by\cite{boumal2023intromanifolds}}
\begin{equation}\label{eq:RIS_Opt_Prob1}
\boldsymbol{\Theta}^{*}=\arg\;\min_{\boldsymbol{\Theta}\in\mathcal{M}}\; f\left(\boldsymbol{\Theta}\right).
\end{equation}
Now, \eqref{eq:RIS_Opt_Prob1} can be solved  by extending the conjugate gradient (CG) method applicable on the Euclidean space to the manifold space with some necessary projections. {To begin with, we  calculate the Euclidean gradient of $f\left(\boldsymbol{\Theta}\right)$ by writing the argument of the summations
in \eqref{eq:Obj_RIS}, which is real-valued, as
\begin{equation}\label{eq:Eu_grad_s1}
\widetilde{f}\left(\boldsymbol{\Theta}\right)=\langle \boldsymbol{\Theta},\mathbf{G}\boldsymbol{\Theta}\mathbf{R}_{j}\boldsymbol{\Theta}^{H}\mathbf{G}\boldsymbol{\Theta}\mathbf{R}_{k}\rangle,
\end{equation}
where $\langle \cdot , \cdot \rangle$ is the inner product. Next we exploit the usual product rules to find the directional derivative $\textrm{D}\widetilde{f}\left(\boldsymbol{\Theta}\right)[\mathbf{V}]$, i.e., variation of $\widetilde{f}$ when we move away from $\boldsymbol{\Theta}$ along the direction $\mathbf{V}$, as\cite[Sec. $4.7$]{boumal2023intromanifolds}
\begin{align}\label{eq:Eu_grad_s2}
    \textrm{D}\widetilde{f}\left(\boldsymbol{\Theta}\right)[\mathbf{V}]& = \langle \mathbf{V},\mathbf{G}\boldsymbol{\Theta}\mathbf{R}_{j}\boldsymbol{\Theta}^{H}\mathbf{G}\boldsymbol{\Theta}\mathbf{R}_{k}\rangle +
    \langle \boldsymbol{\Theta},\mathbf{G}\mathbf{V}\mathbf{R}_{j}\boldsymbol{\Theta}^{H}\mathbf{G}\boldsymbol{\Theta}\mathbf{R}_{k}  \nonumber\\
    & + \mathbf{G}\boldsymbol{\Theta}\mathbf{R}_{j}\mathbf{V}^{H}\mathbf{G}\boldsymbol{\Theta}\mathbf{R}_{k} + \mathbf{G}\boldsymbol{\Theta}\mathbf{R}_{j}\boldsymbol{\Theta}^{H}\mathbf{G}\mathbf{V}\mathbf{R}_{k}  \rangle .
\end{align}
Then, we utilize the properties in \cite[eq. $3.18$] {boumal2023intromanifolds} to rearrange the above expression such that we obtain $\textrm{D}\widetilde{f}\left(\boldsymbol{\Theta}\right)[\mathbf{V}]=\langle \mathbf{V},\nabla\,\widetilde{f}\left(\boldsymbol{\Theta}\right)\rangle$, where $\nabla\,\widetilde{f}\left(\boldsymbol{\Theta}\right)$ is the desired gradient. Following, the Euclidean gradient of $f\left(\boldsymbol{\Theta}\right)$ is written as
\begin{equation}\label{eq:Eu_grad}
\nabla\,f\left(\boldsymbol{\Theta}\right)=\sum_{k=1}^{K-1}\sum_{j=k+1}^{K}2\left(\mathbf{Y}\mathbf{R}_{j}\mathbf{X}\mathbf{R}_{k}+\mathbf{Y}\mathbf{R}_{k}\mathbf{X}\mathbf{R}_{j}\right),
\end{equation}
with $\mathbf{Y}=\mathbf{G}\boldsymbol{\Theta}$ and $\mathbf{X}=\boldsymbol{\Theta}^{H}\mathbf{Y}$. Next, we define the tangent space at a point on $\mathcal{M}$ as $\mathcal{T}_{\boldsymbol{\Theta}}=\{\mathbf{T}\in\mathbb{C}^{N\times N}\mid \mathfrak{R}\{\boldsymbol{\Theta}^{H}\mathbf{T}\}=\mathbf{0}_{N}\}$ which comprises of all the tangent vectors indicating all possible directions this point can move to. By projecting the Euclidean gradient in \eqref{eq:Eu_grad} onto $\mathcal{T}_{\boldsymbol{\Theta}}$, we obtain the \textit{Riemannian} \textit{gradient}, which points towards the steepest descent direction of $f\left(\boldsymbol{\Theta}\right)$}, given by\cite{boumal2023intromanifolds,Hongyu@BDRIS}
\begin{equation}
\begin{split}
\nabla_{\mathcal{M}} f\left(\boldsymbol{\Theta}\right)=&\,\textrm{Pr}_{\boldsymbol{\Theta}}\left(\nabla f\left(\boldsymbol{\Theta}\right)\right)\\
=&\,\nabla f\left(\boldsymbol{\Theta}\right)-\boldsymbol{\Theta}\,\textrm{chdiag}\left(\boldsymbol{\Theta}^{H}\nabla f\left(\boldsymbol{\Theta}\right)\right), 
\end{split}
\end{equation}
where $\textrm{Pr}_{\boldsymbol{\Theta}}(\cdot)$ denotes the projection function, and $\textrm{chdiag}(\cdot)$ constructs a diagonal matrix out of the diagonal elements of the argument matrix. Now, we can apply the CG method in which at the $i^{th}$ iteration, two steps are successively followed
\par {\textit{Step $1$}: determine the descent direction in $\mathcal{T}_{\boldsymbol{\Theta}}$ as}
\begin{equation}\label{eq:step_1}
  {\xi}_{i}=-\nabla_{\mathcal{M}} f\left(\boldsymbol{\Theta}_{i} \right)+\mu_{i}\,\textrm{Pr}_{\boldsymbol{\Theta}_{i}}\left( {\xi}_{i-1}\right),
\end{equation}
where $\mu_{i}$ is the CG update parameter calculated by adopting the Riemannian version of the Polak-Ribi\`ere formula  as\cite{boumal2023intromanifolds}
\begin{equation}\label{eq:step_11}
\mu_{i}=\frac{\mathfrak{R}\left\{\textrm{tr}\left(\mathbf{Z}_{i}^{H}\left(\mathbf{Z}_{i}-\textrm{Pr}_{\boldsymbol{\Theta}_{i}}\left(\mathbf{Z}_{i-1}\right)\right)\right)\right\}}{\textrm{tr}\left(\mathbf{Z}_{i-1}^{H}\mathbf{Z}_{i-1}\right)},
\end{equation}
with $\mathbf{Z}_{i}=\nabla_{\mathcal{M}} f\left(\boldsymbol{\Theta}_{i}\right)$.
\par {\textit{Step $2$}: perform a retraction back to $\mathcal{M}$ as\cite{boumal2023intromanifolds,Hongyu@BDRIS}}
\begin{equation}\label{eq:Theta_i}
\boldsymbol{\Theta}_{i+1}=\left(\boldsymbol{\Theta}_{i}+\delta_{i}\xi_{i}\right)\left(\mathbf{I}_{N}+\delta_{i}^{2}\xi_{i}^{H}\xi_{i}\right)^{-1/2},
\end{equation}
where $\delta_{i}$ is the step size obtained using a back-tracking algorithm\cite{boumal2023intromanifolds}. A local optimal solution of problem \eqref{eq:RIS_Opt_Prob1} can be obtained by appropriately initializing $\boldsymbol{\Theta}$ and then iteratively updating $\xi_{i},\,\mu_{i},\,\boldsymbol{\Theta}_{i}$ and $\delta_{i}$ until convergence. 
\begin{algorithm}[!b]
\caption{Manifold Algorithm}\label{alg:RIS_Opt}
\begin{algorithmic}[1]
\State $\mathbf{Initialize}\:\: i\leftarrow 0,\: \boldsymbol{\Theta}_{0},\,\xi_{0}$
\State \textbf{while} no convergence of $\norm{\nabla_{\mathcal{M}} f\left(\boldsymbol{\Theta}_{i} \right)}_{F}$
\State $\;\;\;\;i\leftarrow i+1;$
\State \;\;\;\;Calculate $\delta_{i}$ by backtracking algorithms \cite{boumal2023intromanifolds};
\State \;\;\;\;Update $\boldsymbol{\Theta}_{i}$ using \eqref{eq:Theta_i}, $\mu_{i}$ using \eqref{eq:step_11}, and $\xi_{i}$ using \eqref{eq:step_1};
\State \textbf{end}
\State  Obtain $\boldsymbol{\Theta}^{*}=\boldsymbol{\Theta}_{i+1}$
\end{algorithmic}
\end{algorithm}
The convergence of manifold algorithm employed to solve constraint optimization problems for BD-RIS architectures is discussed extensively in \cite{Hongyu@BDRIS}. In regard to initialization, we adopt a conventional diagonal matrix in which each diagonal element has unit amplitude and  random phase shift within the range $[0,2\pi)$. Algorithm \ref{alg:RIS_Opt} summarizes the procedure of the manifold algorithm for RIS matrix optimization. Since D-RIS architecture is a special case of BD-RIS architecture, Algorithm \ref{alg:RIS_Opt} can be used for the former as well by evaluating the Euclidean gradient accordingly.
\vspace{-0.1cm}
\subsection{Power control coefficients optimization}
We now proceed to formulate the power control coefficients optimization problem with the aim of maximizing the SE fairness among users (MaxMin) as
\begin{subequations}\label{eq:MaxMin_1}
\begin{align}
\max_{\rho_{1},\ldots,\rho_{K}}&\;\min_{1,\ldots,K}\textrm{SE}_{k}^{\textrm{LB}} \label{eq:MaxMin_1_obj}\\
\textrm{s.t.} &\; \rho_{1}+\ldots+\rho_{K}\leq \rho_{\textrm{dL}},\label{eq:MaxMin_1b}\\
&\; \rho_{k} > 0,\forall k. \label{eq:MaxMin_1c}
\end{align}
\end{subequations}
Since $\log_{2}(\cdot)$ is monotonous, $\textrm{SE}_{k}^{\textrm{LB}}$ can be replaced by  $\gamma_{k}^{\textrm{LB}}$ in \eqref{eq:MaxMin_1_obj}. Problem \eqref{eq:MaxMin_1} can then be equivalently transformed as
\begin{subequations}\label{eq:MaxMin_2}
\begin{align}
\max_{\rho_{1},\ldots,\rho_{K},\gamma}&\;\;\;\;\; \gamma \\
\textrm{s.t.} &\;\frac{\rho_{k}a_{k}}{\sum_{i=1}^{K}\rho_{i}b_{ki}+\sigma_{\textrm{dL}}^{2}}\geq \gamma, \forall k,\\
&\; \rho_{1}+\ldots+\rho_{K}\leq \rho_{\textrm{dL}},\\
&\; \rho_{k} > 0,\forall k, 
\end{align}
\end{subequations}
where $a_{k}$ and $b_{ki}$ can be easily discerned and taken from \eqref{eq:Hardening_B}. Problem \eqref{eq:MaxMin_2} is a standard problem in  MaMIMO literature and can be solved using either the bisection method or the successive convex approximation (SCA) approach employed in previous works\cite{massivemimobook,Anup@MaMIMO}. We briefly detail the procedure to solve \eqref{eq:MaxMin_2} using the Bisection method in Algorithm \ref{alg:Bi_opt}, which solves a sequence of linear feasibility problems until convergence is reached with tolerance $\nu>0$.
\begin{algorithm}[!h]
\caption{Bisection Algorithm}\label{alg:Bi_opt}
\begin{algorithmic}[1]
\State \textbf{Choose}\:\: $\gamma_{\textrm{max}}$ and  $\gamma_{\textrm{min}}$
\State \textbf{while} $\gamma_{\textrm{max}}-\gamma_{\textrm{min}}>\nu$
\State \;\;\;\;Set $\gamma \leftarrow \frac{\gamma_{\textrm{max}}+\gamma_{\textrm{min}}}{2}$
\State \;\;\;\;Solve the following convex feasibility problem 
\item[]\;\;\;\;
 $\begin{cases}
\,\,\gamma\left(\sum_{i=1}^{K}\rho_{i}b_{ki}+\sum_{\textrm{dL}}^{2}\right)-\rho_{k}a_{k}\leq 0,\,\forall k,\\
   \,\,\rho_{1}+\ldots+\rho_{K}\leq \rho_{\textrm{dL}}.
  \end{cases}$
\State\;\;\;\;\textbf{if} \textit{feasible} \textbf{then}
\item[]\;\;\;\;\;\; $\gamma_{min}\leftarrow \gamma$, 
 $\rho_{k}^{*}\leftarrow \rho_{k},\,\forall k$ 
\State\;\;\;\;\textbf{else}
\item[]\;\;\;\;\;\; $\gamma_{max}\leftarrow \gamma$ 
\State \textbf{end}
\State \textbf{Obtain} $\rho_{1}^{*},\ldots,\rho_{K}^{*}$
\end{algorithmic}
\end{algorithm}

\section{Numerical Results}\label{NumRes}
 To assess the SE performance of BD-RIS directed MaMIMO transmission (MaMIMO BD-RIS), numerical results are illustrated for the topology depicted in Fig.~\ref{fig:BS_Side_RIS}. The users are randomly distributed within an angular sector $[-\pi/3,\pi/3]$ between the distances $[10,\,400]$ m from the RIS. The path-loss for the RIS-user channel is modelled according to \cite[eq. 2.3]{massivemimobook}. The network parameters are reported in Table~\ref{tab:Parameter_table}. For comparison, SE performance of a conventional TDD MaMIMO transmission (MaMIMO) is considered with $\tau_{up}=K$ and Kronecker $3$D channel model between the BS and user, for the same given user topology.
\begin{table}[!t]
	\caption{Simulation parameters}\vspace{-0.2cm}
	    \label{tab:Parameter_table}\centering
	\begin{tabular}{l l}
		\toprule[0.4mm]
		\textbf{Parameter}&\textbf{Value}\\
   {{Carrier frequency}, {bandwidth}} & {{$f_{c}=2.5\,\textrm{GHz}$}, $B=20\,\textrm{MHz}$}\\
   {RIS and UE heights} & $h_{\textrm{RIS}} = 10\,\textrm{m}$, $h_{\textrm{u}} = 1.5\,\textrm{m}$\\
   Parameters of BS-RIS channel\cite{buzzi2022approaching} & $\varrho=1,\,G_{A}\left(\theta\right)=G_{R}\left(\theta\right)=3\textrm{dB}$ \\
   TDD parameters (samples) & {$\tau = 200,\,\tau_{p}=K,\,\tau_{up}=QK$}\\
   Total transmit powers &  $\rho_{\textrm{uL}} = 400\,\textrm{mW}$, $\rho_{\textrm{dL}}= 1200\,\textrm{mW}$\\
   Noise powers \cite{massivemimobook} & $\sigma_{\textrm{uL}}^{2}=\sigma_{\textrm{dL}}^{2}=-94\,\textrm{dBm}$\\
		\bottomrule[0.4mm]
	\end{tabular}\vspace{-0.4cm}
\end{table}   
\par
\begin{figure}[!t]
    \centering
\includegraphics[width=0.5\columnwidth,height=6cm,tics=10]{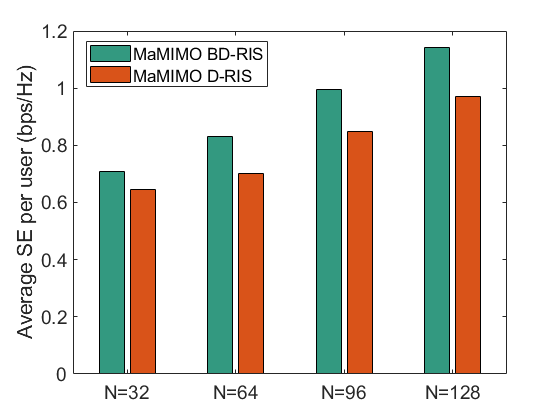}
    \caption{Average SE per user performance of MaMIMO BD-RIS and MaMIMO D-RIS for $M=24$ and $K=4$.}
    \label{fig:FCVsSC}\vspace{-0.7cm}
\end{figure}

 In Fig.~\ref{fig:FCVsSC}, average SE per user performance of MaMIMO BD-RIS with respect to $N$ is illustrated and compared with that of D-RIS directed transmission (MaMIMO D-RIS). The SE performance of both increase as $N$ increases by virtue of the channel `enhancement' offered by the RIS. Moreover, the relative SE gain of MaMIMO BD-RIS over MaMIMO D-RIS increases from $8.72\%$ for $N=32$ to $15.45\%$ for $N=64$. However, the gain drops to $15.01\%$ for $N=128$, implying eventual saturation of the relative gain with increasing $N$. Nevertheless, the advantage of using BD-RIS is significant owing its superior fully-connected architecture, and ability to adjust both magnitude and phase of the impinging waves.
\par
\begin{figure}
    \centering
\subfloat[ $K=8$]{\includegraphics[width=0.52\columnwidth, height=6cm,tics=10]{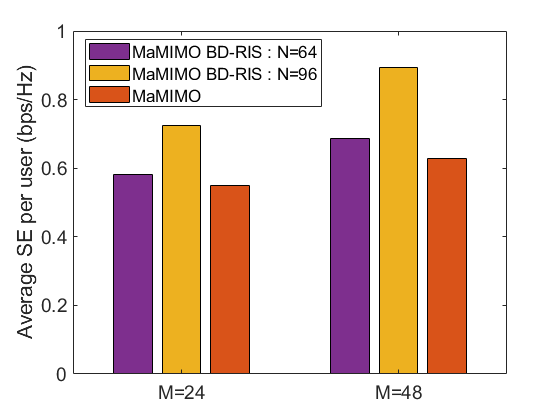}}
\subfloat[$M=32$ and $N=128$]{\includegraphics[width=0.52\columnwidth,height=6cm,tics=10]{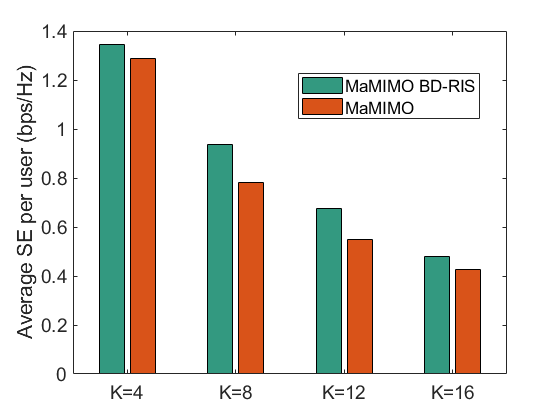}}
    \caption{Average SE per user MaMIMO BD-RIS vs MaMIMO.}
    \label{fig:FCVsNVsM}\vspace{-0.6cm}
\end{figure}

 {Next, we illustrate the SE comparison of MaMIMO BD-RIS and MaMIMO in Fig.~\ref{fig:FCVsNVsM}. To ensure fairness in degrees of freedom (DoF) available for comparison, we keep $M$ same for both and $N_{H}=M_{H}\geq K$}. Note that the channel hardens less in the presence of RIS\cite{Channel@Hardening}. Therefore, a higher $N$ is required when $M$ is small to offset the inimical effect and achieve a significant SE gain over MaMIMO. This is validated in Fig.~\ref{fig:FCVsNVsM}(a) where for $M=24$, the SE gain over MaMIMO increases from $5.5\%$ for $N=64$ to $24\%$ for $N=96$. Moreover, with increasing $M$, channel hardens more and allows for better exploitation of the RIS. As a result, for $N=96$, SE gain of MaMIMO BD-RIS increases from $24\%$ for $M=24$ to $30\%$ for $M=48$. Imperatively, MaMIMO BD-RIS with $M=24,\,N=96$ achieves a $13\%$ higher SE than MaMIMO with $M=48$, thereby exhibiting its capability to reduce the energy expensive $M$ to achieve a specific SE.
\par In Fig.~\ref{fig:FCVsNVsM}(b), to illustrate and compare the SE performance with respect to $K$ we choose $M=32$ and $N=128$ such that $M_{H}=N_{H}=16$. Since we have MaxMin for power control coefficients optimization, naturally, as $K$ increases the SE decreases. But, the SE gain of MaMIMO BD-RIS over MaMIMO increases from $4.4\%$ for $K=4$ to $19.09\%$ for $K=12$. It is because if $K\ll{M}_{H}$ then MaMIMO itself has sufficient DoF to achieve a good SE and the gain of MaMIMO BD-RIS is very less. However, as $K$ increases, benefits of large DoF wanes for MaMIMO, whereas the ability to tune RIS elements allows us to force a larger degree of orthogonality between composite channels of users providing additional DoF for MaMIMO BD-RIS to serve more users with a good SE performance. Finally, as $K$ increases further to $16$, this SE gain drops to $11\%$ due to the excessive increase in $\tau_{up}$, which decreases the pre-log factor in \eqref{eq:SE_Exp} and lowers the SE of MaMIMO BD-RIS. Therefore, depending on $M$ and $N$, MaMIMO BD-RIS SE gain over MaMIMO will first increase with $K$, but as $K$ continues to increase, the gain will decrease due to the channel estimation overhead, and eventually MaMIMO BD-RIS will loose its superiority.    
\vspace{-0.25cm}
\section{Conclusion}\label{Concl}
In this letter, we considered a BS side BD-RIS deployment and proposed a transmission framework for a MaMIMO network. Based on statistical CSI, we optimized the BD-RIS matrix and power control coefficients using a novel manifold algorithm and convex optimization, respectively. Through numerical results we first demonstrated the superiority of the BD-RIS architecture over D-RIS in terms of SE. Thereafter, we illustrated  different network settings for which the framework outperforms a conventional MaMIMO transmission.

\ifCLASSOPTIONcaptionsoff
  \newpage
\fi

\appendices
\vspace{-0.2cm}
\bibliographystyle{IEEEtran}
\bibliography{reference}
\end{document}